 \documentclass[preprint2]{aastex}

\shorttitle{Super Li-rich star}
\shortauthors{Laws \& Gonzalez}

\begin{document}

\title{A Re-evaluation of the Super Li-rich Star in NGC 6633}


\author{Chris Laws}
\affil{Astronomy Department, University of Washington, Seattle, WA 98195}
\author{Guillermo Gonzalez}
\affil{Department of Physics and Astronomy, Iowa State University, 
Ames, IA 50011}

\begin{abstract}

We present a new abundance analysis of the super Li-rich star J37 and a 
comparison star in NGC 6633. We confirm the result of Deliyannis et al. 
that J37 has a Li abundance well above the meteoritic value, and we also 
confirm that Al, S, Si, Ca, Fe and Ni are super-solar, and C is sub-solar.
We additionally find that Na, Sc, Ti are super-solar, while O is sub-solar.
The abundance pattern of metals in J37 is generally consistent
with the bulk composition of the Earth. We propose that accretion of
circumstellar matter is the best explanation for these abundance anomalies,
although we cannot rule out a secondary contribution from diffusion.

\end{abstract}


\section{Introduction}

Deliyannis, Steinhauer, \& Jeffries (2002, hereafter DSJ02) reported on 
their detailed abundance spectroscopic analysis of the star J37 in the open 
cluster NGC 6633. NGC 6633 is comparable in age to the Hyades, but it is 
slightly metal-poor. DSJ02 discovered that J37 has a Li abundance 
near $A(Li) = 4.3~[A(X) = 12 + \log (N_{X}/N_{H})]$, one dex above the 
Solar System meteoritic value. They determined abundances for eight 
elements, all of which were found to be enhanced above solar except C.

DSJ02 could not account for all the abundance anomalies in J37 with any 
known process. They favored diffusion as the best overall explanation and 
concluded that J37 is the first member of a previously suggested ``Li 
peak'', covering a narrow temperature window between 6900 and 7100 K. 
However, they admitted that recent stellar diffusion models do not show 
a Li peak. DSJ02 also dismissed planetesimal accretion as a plausible 
explanation, owing to a poor match between the observed abundance anomalies
and expectations from condensation temperature arguments -- although
similar concerns plague the diffusion model interpretation. We will revisit
both explanations in the present study.

In order to reach a better understanding of the nature of the abundance 
anomalies in J37, we obtained spectra of J37 and the comparison star J1 
in the same cluster; DSJ02 had employed a comparison star in the open 
cluster M35. In the following we report on the results of our abundance 
analyses for both stars based on the new data.

\section{Observations}

Since the purpose of our study is to determine the quantitative abundance 
differences between J37 and its host cluster, it is necessary to select a 
suitable comparison star. Based on its location on the color-magnitude (CM) 
diagram of NGC 6633, its high probability of membership, and similar $v 
\sin i$, we selected J1 as the best comparison star for J37 (Jeffries et 
al. 2002). Unfortunately, Jeffries et al. list J1 as a probable binary 
based on its location on the CM diagram relative to the ZAMS. The results 
of our abundance analysis of J1 (presented below) yields a metallicity close 
to the cluster mean reported by Jeffries et al., so it is unlikely that a 
possible companion to J1 has significantly compromised our analysis of it. We 
also observed J2, but subsequent analysis showed that our data were not of 
sufficient quality to complete a reliable analysis of this star.

We obtained optical spectra of NGC 6633 members J1, J2, and J37 over a span
of approximately three hours on October 11, 2002, with the Apache Point
Observatory 3.5m telescope. We employed the ARC Echelle Spectrograph and 
a 2048x2048 SiTe CCD, and made no changes to the instrument between the 
two observations. We achieved a resolving power of $\sim$ 37,000 (as 
measured on a Th-Ar lamp spectrum) with a signal-to-noise ratio (S/N) of 
$\sim$200 per pixel at 6700 \AA. Spectra of a hot star were also
obtained at a similar airmass in order to compensate for telluric features
in the spectra of J1 and J37.

Data for both stars were reduced in as nearly an identical manner as 
possible, following the procedures described in the University of Chicago's
IRAF Data Reduction Guide for the ARC Echelle Spectrograph (prepared by Julie
Thompson\footnote{$http://www.apo.nmsu.edu/Instruments/echelle/ARCES\_data\_guide.pdf$}).
Each image was processed with the same bias and flat fields, and 
corrections for scattered light were made with the same fitting functions 
in both spectra. In addition, continua of individual orders were normalized 
with exactly the same functions in both stars. These steps were undertaken 
to minimize possible systematic differences between line depths which would 
ultimately impact our differential analysis. Our efforts yielded continuous 
one-dimensional spectra in the blue up to 4000 \AA, with gaps between orders 
thereafter up to about 10,000 \AA.

\section{Analysis}

The reduced spectra were examined carefully for suitable, largely unblended
atomic lines. The equivalent widths (EWs) of these lines were measured following
standard procedures within IRAF. Table 1 presents our adopted linelist and
measured EWs for J1 and J37. Because of the relatively high stellar rotation
velocities compared to cooler dwarfs, we could not employ many of the lines
we have used in our previous studies of solar-type stars. We adopted some 
gf-values from Gonzalez \& Wallerstein (1999), and some are from accurate 
laboratory estimates available in the literature.

\subsection{Atmospheric Parameters}

We employed the measured Fe I and Fe II EWs and associated $gf$-values from 
Table 1 to derive a set of basic stellar parameters ($T_{\rm eff}$, $\log g$,
$\xi_{\rm t}$, and [Fe/H]) and associated uncertainties for J1 and J37. All
analyses utilized a recent version of the LTE abundance code MOOG (Sneden, 1973)
and Kurucz model atmospheres without convective overshoot (Castelli et al. 1997
\footnote{All model atmospheres used are from $http://kurucz.harvard.edu/grids.html$.}).
An initial set of model atmospheres was selected with a range of values of $T_{\rm eff}$,
$\log g$, $\xi_{\rm t}$, and [Fe/H] centered on the results of DSJ02. Values for
$\log g$ and $\xi_{\rm t}$ were then determined by systematically iterating all
four of the basic parameters until the mean Fe I and Fe II abundances were equal
and correlations of the individual Fe I line abundances with both $\chi_{\rm l}$
and the logarithm of reduced equivalent width (REW) were zero.

This procedure additionally yielded spectroscopic estimates of $T_{\rm 
eff}(spec)=7305\pm242$ for J37 and $6950\pm223$ for J1. The relatively large 
uncertainties in these values are due to the small number of quality Fe I 
lines available -- especially those of low excitation potential -- as well as
to known non-LTE effects on Fe I abundances. The resulting uncertainty in $T_{\rm eff}$
is the most important source of uncertainty in subsequently derived abundances.
To ameliorate this, we adopted a weighted average of our spectroscopic $T_{\rm eff}$
values and estimates based on photometry, using the equation presented in DSJ02
and the $B-V$ and $E(B-V)$ values quoted by Jeffries (1997). An additional possible
concern is the fact that the abundance mix of J37 is significantly different from solar
and that this will affect the accuracy of the photometric $T_{\rm eff}$ estimate.

This weighted average $T_{\rm eff}$ was then used with the spectroscopic
$\log g$ to determine [Fe/H] from linear interpolation of the MOOG Fe I and
Fe II abundance results. Finally, a new set of model atmospheres were selected
centered on these $T_{\rm eff}$(avg), $\log g$, $\xi_{\rm t}$, and [Fe/H] values
and the entire process iterated in metallicity until it converged.

A summary of the results of the above procedure is given in Table 2, along 
with estimates of absolute V-band magnitudes (M$_{\rm V}$), theoretical 
values of $\log g$, and mass estimates; [Fe/H] values are given in Table 1.
Theoretical estimates of $\log g$, and mass are derived from our adopted
$T_{\rm eff}$(avg), M$_{\rm V}$, and a set of 600 Myr Padova Stellar 
Isochrones (Jeffries et al. 2002; Salasnich et al. 2000) appropriate to each 
star's metallicity.

As noted above, Fe I abundances are known to suffer from non-LTE effects. These
do not affect Fe II abundances, and this difference could lead to underestimates
in our spectroscopic $\log g$. We find, however, that our theoretical and spectroscopic
surface gravity values are in good agreement for J1. A star of similar $T_{\rm eff}$
and $\log g$, HE 490 in the $\alpha$ Per cluster, was studied in Gonzalez \& Lambert (1996)
and they also found no significant offset between theoretical and spectroscopic values
of $\log g$ to within their uncertainties of $\sim$ 0.1 dex. In the case of J37,
however these two differ considerably -- much more so than non-LTE effects seem
capable of producing. We offer a different possible explanation of this discrepancy below.

\subsection{Additional Elements}

Abundances for C, O, Na, Mg, Si, S, Ca, Sc, Ti, and Ni were determined in 
the same manner as for [Fe/H], employing our measured EWs and our values of 
$T_{\rm eff}$(avg) and $\log g$(spec) to interpolate the abundances from MOOG 
outputs. Table 1 summarizes these results, with associated uncertainties
estimated from the mean deviation of line-by-line results and linear 
propagation of uncertainties in $T_{\rm eff}$(avg) and $\log g$(spec).
The quoted differential O abundance for J1 and J37 contains an -0.57 dex offset
to account for non-LTE effects on the 7771-5 O I triplet (Takeda, 1997). The
differential Na abundance has also been adjusted by -0.2 dex for non-LTE using
offsets from lines of similar EWs reported in Bikmaev et al. (2002). 

We determined abundances of Li and Al through comparison of the observed
spectra between 6700 and 6715 \AA\ and synthesized spectra. The latter were
produced with the same set of stellar models utilized in determining [Fe/H]
in each star and a linelist of the region we have used in previous studies
(c.f. Reddy et al. 2002). The results are included in Table 1, with 
uncertainties estimated by visual inspection of synthesized spectra within 
an appropriate range of model conditions and elemental abundances.

Finally, we present in Table 1 $\Delta$[X/H] values, calculated as the mean
deviation of the differences in abundance values reported on a line-by-line
basis between J1 and J37 \footnote{$\Delta$[Li/H] and $\Delta$[Al/H] Li are
exceptions to this, and are simply the differences between the reported
abundances of the spectral syntheses)}. As discussed in Laws \& Gonzalez
(2001), such line-by-line differential comparisons eliminate systematic errors
caused by uncertain $gf$-values and can provide a significant increase in 
precision over differencing the average abundances alone.

\section{Discussion}

\subsection{Comparisons}

Our abundance results for J37 are consistent with those reported by DSJ02, 
confirming the anomalous nature of this star. In particular, we verify that 
lithium is enhanced by about a factor of 10 relative to the Solar System 
meteoritic value. In addition, iron is about half a dex higher than the 
solar value, while carbon is strongly depleted. We do find a lower sulfur 
abundance than DSJ02, but their estimate for this element was based on only 
one line.

\subsection{Possible causes of abundance pattern}

J37 is a late A or early F dwarf, and as such, it should possess a very
shallow convection zone. This relatively thin mixing layer allows certain
processes to produce observable effects that are mostly negligible in cooler
dwarfs. Two that have been discussed in the literature are accretion and
diffusion. DSJ02 favored diffusion as the best explanation for the abundance 
pattern in J37. In light of the new observations we present in this study, we 
revisit these two hypotheses below.

\subsubsection{Diffusion}

Richer, Michaud, \& Turcotte (2000, hereafter RMT00) present models that 
include the effects of atomic and turbulent diffusion transport in an effort 
to account for the abundance trends observed in AmFm stars. They are 
successful in accounting for the general abundance patterns seen, especially 
the increase in abundance with atomic number. There do still exist some significant 
discrepancies between the models and observations, but it is not clear how
much can be attributed to observational versus theoretical uncertainties.

DSJ02 argued that the diffusion model abundances of RMT00 offer the best 
explanation for the anomalous abundance pattern of J37 in relation to a 
comparison star in M35. DSJ02 note that the theoretical results of RMT00 are 
consistent with their abundance estimates for C, Fe, and Ni. However, 
they also note that their abundance estimates for S, Si, Al, Ca, and 
especially Li (but see below) are too high compared to the predictions of 
RMT00.

We added four elements not measured by DSJ02 in J37: O, Na, Sc, and Ti. 
While our LTE analysis of the 7771-5 O triplet showed very similar oxygen
abundances for J1 and J37, the non-LTE corrections for abundances derived
from these features (Takeda, 1997) yield a considerable differential offset
between these two stars, resulting in J37 appearing depleted in O by 0.57 dex
relative to J1. Na, Sc, and Ti, however are significantly enhanced relative to J1.
RMT00 predict O to be depleted, Sc and Ti to be slightly enhanced, and Na to be 
slightly depleted. Our lower abundance value for S than DSJ02 mitigates the 
disagreement with RMT00, but it is still too high. In summary, then, the 
predictions of RMT00 are not in agreement with the observed abundances of the 
elements Li, Na, S, Si, Al, Ca, and Sc, while they are consistent 
with C, O, Fe, and Ni (and possibly Ti). The very low abundance of C and the
non-LTE corrected O abundance are the strongest evidence for diffusion in J37.
We compare in Figure 1 our abundance results to the abundances of 63 Tau and
the best-fit diffusion model from RMT00; 63 Tau is an A star in the Hyades cluster
with a similar T$_{\rm eff}$ to J37.

DSJ02 suggested the high abundance of Li could be explained by the 
theoretical model of Richer \& Michaud (1993). However, they also 
noted that the more recent models of RMT00 no longer showed enhancement 
of Li. The Li enhancement predicted by Richer \& Michaud displays a narrow 
peak centered just below 7100 K for [Fe/H] = $-0.15$. DSJ02 find that J37 
falls on the hot side of the predicted Li-peak. Our slightly higher 
T$_{\rm eff}$ for J37 puts it even farther from the predicted peak. At the 
same time, our analysis places J1 on the cool side of the predicted Li-peak, 
yet it displays no apparent Li enhancement.

\subsubsection{Accretion}

Alexander (1967) first proposed that the surface composition of a star could 
be measurably altered by the accretion of H-depleted material; specifically, 
he argued that an old star could experience dramatic enhancement of its Li 
abundance with the accretion of a planet like Jupiter. Brown et al. (1989) 
revived the idea to try to account for the very high Li abundances observed in 
some field red giant stars. Gonzalez (1998) suggested that planet accretion 
might account for unusual observed stellar atmospheric abundances -- this time
for Sun-like stars with giant planets.

Laws \& Gonzalez (2001) and Gratton et al. (2001) present evidence for the
alteration of the surface abundances of stars in wide binaries (16 Cyg 
and HD 219542, respectively) by accretion of H-depleted material. The 
components of 16 Cyg and HD 219542 (all G dwarfs) display slightly different 
metallicities, and even different trends with condensation temperature, a 
strong sign of accretion of fractionated material. The sense of the trend 
is such that elements with high condensation temperature have higher 
abundances compared to most stars. The abundances are enhanced by less than 
a tenth of a dex in these cases.

The interstellar medium gas phase abundances correlate strongly with 
condensation temperature. This is interpreted as resulting from
preferential condensation onto grains by elements with high condensation 
temperature. Two classes of stars are observed to display abundances that 
correlate with the interstellar abundances: post-AGB and $\lambda$ Boo 
stars. The former are in a highly evolved stage, while the latter are 
young (late B to early F dwarfs). The post-AGB class includes variables 
(Giridhar et al. 2000) and non-variables (van Winckel et al. 1992). For 
example, Giridhar et al. find that those RV Tauri variables showing evidence 
of depletion often have Al/Fe, Ca/Fe, and Sc/Fe ratios below solar, while 
C/Fe, O/Fe, S/Fe, and Na/Fe are above solar. Because populations I and II are 
present among the post-AGB stars, it is not always clear what the original 
abundances were prior to selective depletion. It is usually assumed that the 
original abundance scale is given by their S and Zn abundances.

$\lambda$ Boo stars display above-solar ratios for C/Fe, N/Fe, O/Fe, S/Fe 
and below-solar ratios for Ca/Fe and Ti/Fe. Venn \& Lambert (1990) proposed 
that the abundance pattern of $\lambda$ Boo stars derive from their local 
circumstellar or interstellar environment, given their correlation with 
the interstellar gas phase abundances. Heiter (2002) reviews the abundance 
patterns of $\lambda$ Boo stars using all available data published through 
2001. Interestingly, they find that Na is often enhanced above solar, and 
the Ni/Fe ratio is also enhanced; Andrievsky et al. (2002) confirm the 
high Na abundances in some $\lambda$ Boo stars. Heiter et al. (2002) 
confirmed that the abundances of $\lambda$ Boo stars do correlate with the 
interstellar abundances in general, but they differ in detail. Paunzen et al. 
(1999) have found very low C abundances for some $\lambda$ Boo stars, but 
there is still some controversy concerning C abundances determined from near 
infrared versus optical lines.

Thus, at present there is observational evidence for two processes that 
can alter the surface abundances of stars as a result of accretion of 
fractionated material. In one case, material depleted in low-condensation 
temperature elements accretes onto a star and enhances the abundances of 
elements with high condensation temperature. In the other, gas selectively 
depleted in high condensation temperature elements via grain condensation 
and loss is accreted onto a star, resulting in depletion of the high condensation 
temperature elements in its atmosphere. In the first case, relatively small 
amounts of accreted material can cause observed surface abundance changes in 
a star, while in the second, most of the mass of a star's convective envelope 
must be composed of accreted gas.

\subsubsection{Diffusion and Accretion?}

Can we select one process over the other in the case of J37? Based on the 
above discussion, it appears that neither diffusion nor accretion (of 
either type) is completely consistent with its abundance pattern. Its low C 
and O abundances and above-solar Ni/Fe ratio are consistent with diffusion, but the 
high Li, Na, Mg, Al, Si, S, Ca, Sc, and Ti abundances are not consistent with 
it. Its overall abundance pattern does correlate with condensation temperature, but 
there are some anomalies, such as the high Na abundance and low C/O and 
high Ni/Fe ratios.

Diffusion and accretion processes are observed over the same range in 
spectral type: A-F dwarfs. Both processes are most effective for stars with 
a shallow convection zone. Therefore, it would be reasonable to expect that 
many stars should display the effects of both, though in most cases one or the 
other will dominate. Until now, studies have focused on either diffusion or 
accretion to explain the non-solar abundance patterns among early-type dwarfs. 
Perhaps we have yet to see a completely clean signature of either process 
operating in a given star.

J37 is near the cool end of the A and F stars observed to have abundance 
anomalies. Its rotation is also slower than most A and early F stars 
($v \sin i = 32$ km~s$^{\rm -1}$ according to DSJ02; we find $25$ km~s$^{\rm 
-1}$ in our syntheses). For example, most $\lambda$ Boo stars studied to date
have $v \sin i$ between 50 and 100 km~s$^{\rm -1}$. Slow rotators are expected
to have less meridional circulation, which tends to dilute surface abundance
anomalies. Given this, J37 might display abundance anomalies more clearly than
otherwise similar stars with faster rotation.

We can combine three separate processes with various weights to account for 
the abundances of J37: accretion of planetesimals, accretion of gas depleted 
of refractories, and diffusion. Unfortunately, neither diffusion theory 
proposed for AmFm stars nor accretion scenarios proposed for $\lambda$ Boo 
stars can account for all the abundance anomalies observed in A and F dwarfs. 
In addition to comparing the abundances of J37 to theory, we should also 
compare them to the observed patterns. If we take into account all the 
abundance anomalies noted among the $\lambda$ Boo stars, we could account 
for its low C and high Na abundances and high Ni/Fe ratio. To further 
account for the high iron-peak abundances in J37, we can add either diffusion 
or planetesimal accretion. The most serious problem with diffusion as an 
explanation for J37 is its prediction of low Li, Ca, Sc, and Ti. But, Ca, 
Sc, and Ti are also observed to be depleted in $\lambda$ Boo stars. Thus, 
planetesimal accretion is also needed.

The composition of accreted planetesimals is important. To account for the 
low C abundance in J37, they must be depleted in C. But, at the same time, 
they should not be strongly depleted in Na or S. The bulk Earth fits these 
requirements rather well, except that Na and S are probably a bit too low in 
the Earth. Table 3 presents [X/Fe] values for J37-J1, terrestrial material
(see Gonzalez et al. 2001 for details), mean $\lambda$ Boo (Heiter 2002), and 
both 63 Tau and a best-fit diffusion model for it (RMT00).

Either diffusion or accretion will result in an enhancement of metals
in the surface layers of a star. The result of this is a spectroscopic $\log g$
value which is smaller than predicted from stellar evolutionary models which
assume homogeneous composition (Laws et al. 2003). Our spectroscopic estimate
of surface gravity for J37 is indeed much lower than that predicted by the
Padova isochrone set (see Table 2), and likely lower than a simple non-LTE
correction could account for.

In summary, we suggest that the elements in the convection zone of J37 were 
set by the composition of: (a) the primordial gas from the star's formation,
(b) subsequent accretion of both depleted circumstellar gas and planetesimal 
material, and (c) evolution of its atmosphere through internal processes 
(diffusion, growth of the convective zone, etc.). The timing of each of these 
is not clear at this time, and some of these processes would certainly affect
the others. For example, the accretion of planetesimals would increase the
metallicity of the atmosphere, and, as a result, deepen the convection zone
and reduce the efficiency of diffusion.

\section{Conclusions}

We have obtained and analysed high quality echelle spectra of J37 and J1
in NGC 6633. Our estimates of elemental abundances are consistent with
those reported by DSJ02 for J37, and we additionally analysed four elements
not measured by DSJ02: O, Na, Sc, and Ti. Overall, we find that the abundance
pattern of J37 shows a very poor match to the predictions of recent diffusion
models, and while we do not rule out diffusion as a factor in the abundance
anomalies seen in J37, it appears to play at best a secondary role. We
suggest instead that accretion of circumstellar material provides the best
match. Although no single accretion source seems completely consistent with
the abundance pattern of J37, we note that the accretion of material depleted
in C, like that of the bulk Earth, is implied. Precisely how much condensed
material and/or depleted gas the star would need to accrete, however, and
whether or not that was reasonable given the environment and evolutionary
state of the star, are details that are difficult to determine at this time.
A larger sample of comparison stars and detailed observations of their
circumstellar environments (including that of J37) are needed to constrain
environmental variables, while further developments in stellar models including
accretion and diffusion processes are also necessary in order to fully resolve
the abundances anomalies observed in this unusual star.

\acknowledgements

This work has been supported by the University of Washington Astrobiology
Program and a grant from the National Astrobiology Institute. Additionally,
this research has made use of the Simbad database, operated at CDS, Strasbourg,
France.

\begin{figure}
\plotone{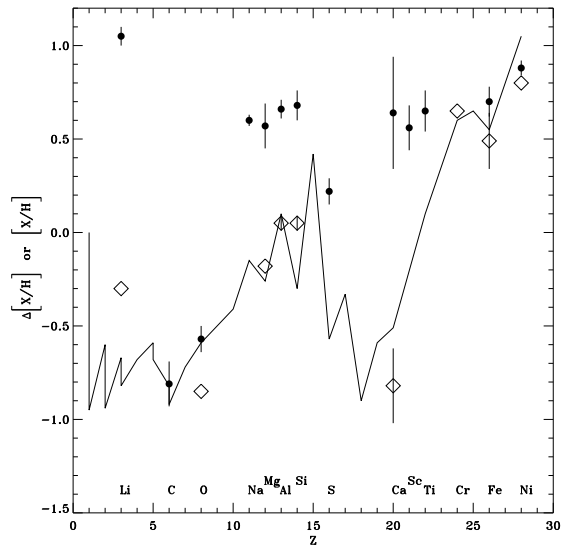}
\caption{$\Delta$[X/H] values for J\,1 and J\,37 (filled circles) from this 
study. Also shown are [X/H] values for 63 Tau (diamonds) and a best-fit model 
that includes diffusion (line), both from RMT(00). Error bars are included for 
those elements in 63 Tau for which uncertainty estimates were provided in RMT(00).}
\end{figure}

\begin{deluxetable}{lcccccc}
\tablecaption{Spectral Results for J\,1 and J\,37}
\scriptsize
\tablewidth{0pt}
\tablehead{
\colhead{Species} & \colhead{$\lambda_{\rm o}$(\AA)} & 
\colhead{EW$_{J\,1}$} & \colhead{EW$_{\rm J\,37}$} &
\colhead{[X/H]$_{\rm J\,1}$} & \colhead{[X/H]$_{\rm J\,37}$} &
\colhead{$\Delta$[X/H](J\,37 - J\,1)}
}
\startdata
C I & 6587.62 & 61.7 & $<$ 31.4 & \nl
C I & 7115.18 & 64.4 & $<$ 37.1 & \nl
C I & 8335.15 & 219.0 & 165.5 & \nl
ave C & \nodata & \nodata & \nodata & $+0.07 
\pm 0.04$ & $-0.74 \pm 0.12$ & $-0.81 \pm 0.12$\nl
O I & 7771.95 & 241.3 & 285.9 & \nl
O I & 7774.18 & 205.1 & 260.0 & \nl
O I & 7775.40 & 177.7 & 216.7 & \nl
ave O & \nodata & \nodata & \nodata & $+0.63 \pm 0.07$ & $+0.58 \pm 0.10$ & $-0.57 \pm 0.07$\tablenotemark{a}\nl
Na I & 8183.25 & 184.2 & 261.0 & \nl
Na I & 8194.84 & 234.9 & 315.0 & \nl
ave Na & \nodata & \nodata & \nodata & $+0.08 \pm 0.01$ & $+0.87 \pm 0.04$ & $+0.60 \pm 0.03$\tablenotemark{b}\nl
Mg I & 8717.83 & 69.9 & 139.3 & \nl
ave Mg & \nodata & \nodata & \nodata & $-0.04 \pm 0.15$ & $+0.53 \pm 0.14$ & $+0.57 \pm 0.12$\nl
Si I & 6145.02 & 17.5 & 57.0 & \nl
Si I & 6721.85 & 30.8 & 64.8 & \nl
Si I & 8556.80 & 138.7 & 235.2 & \nl
ave Si & \nodata & \nodata & \nodata & $-0.13 \pm 0.16$ & $+0.56 \pm 0.19$ & $+0.68 \pm 0.08$\nl
S I & 6046.02 & 31.7 & 68.0 & \nl
S I & 6757.19 & 47.9 & 80.0 & \nl
ave S & \nodata & \nodata & \nodata & $+0.14 \pm 0.11$ & $+0.36 \pm 0.18$ & $+0.22 \pm 0.07$\nl
Ca I & 5857.46 & 121.6 & 211.7 & \nl
Ca I & 6439.08 & 180.5 & 221.6 & \nl
ave Ca & \nodata & \nodata & \nodata & $+0.06 \pm 0.10$ & $+0.71 \pm 0.19$ & $+0.64 \pm 0.30$\nl
Sc II & 6604.60 & 21.1 & 75.2 & \nl
ave Sc & \nodata & \nodata & \nodata & $-0.27 \pm 0.16$ & $+0.30 \pm 0.15$ & $+0.56 \pm 0.12$\nl
Ti II & 5336.77 & 81.9 & 174.8 & \nl
ave Ti & \nodata & \nodata & \nodata & $-0.18 \pm 0.13$ & $+0.47 \pm 0.15$ & $+0.88 \pm 0.11$\nl
Fe I & 5434.52 & 136.9 & 204.7 & \nl
Fe I & 5862.35 & 60.9 & 121.1 & \nl
Fe I & 6027.06 & 36.9 & 81.9 & \nl
Fe I & 6056.01 & 53.8 & 103.0 & \nl
Fe I & 6065.48 & 88.9 & 149.0 & \nl
Fe I & 6380.74 & 18.6 & 50.4 & \nl
Fe I & 6677.99 & 91.6 & 182.4 & \nl
Fe I & 6750.16 & 35.3 & 62.4 & \nl
Fe I & 7507.27 & 32.1 & 78.4 & \nl
Fe I & 7568.91 & 50.8 & 93.1 & \nl
Fe I & 7583.80 & 37.5 & 82.6 & \nl
Fe I & 7586.03 & 75.6 & 143.4 & \nl
Fe I & 8327.05 & 129.8 & 175.7 & \nl
Fe I & 8468.40 & 81.2 & 178.1 & \nl
Fe II & 6084.17 & 30.9 & 82.5 & \nl
Fe II & 6149.25 & 58.0 & 164.5 & \nl
Fe II & 6369.45 & 117.3 & \nodata & \nl
Fe II & 6416.99 & 51.1 & 155.4 & \nl
ave Fe & \nodata & \nodata & \nodata & $-0.19 \pm 0.14$ & $+0.51 \pm 0.16$ & $+0.70 \pm 0.08$\nl
Ni I & 6643.64 & 33.3 & 93.5 & \nl
Ni I & 6767.78 & 26.5 & 73.7 & \nl
ave Ni & \nodata & \nodata & \nodata & $-0.44 \pm 0.08$ & $+0.45 \pm 0.05$ & $+0.88 \pm 0.04$\nl
Li\tablenotemark{b} & \nodata & \nodata & \nodata & $+2.04 \pm 0.05$ & $+3.09 \pm 0.05$ & $+1.05 \pm 0.05$\nl
Al\tablenotemark{b} & \nodata & \nodata & \nodata & $-0.14 \pm 0.05$ & $+0.52 \pm 0.04$ & $+0.66 \pm 0.05$\nl
\enddata
\tablenotetext{a}{An additional -0.52 dex offset has been added to the differential O abundance to account for non-LTE 
effects (Takeda, 1997).}
\tablenotetext{b}{An additional -0.20 dex offset has been added to the differential Na abundance to account for 
non-LTE effects (Bikmaev, et al. 2002).}
\tablenotetext{c}{Abundances and uncertainties estimated from comparisons of observed and synthetic spectra.}
\end{deluxetable}

\begin{deluxetable}{lccccccccc}
\tablecaption{Stellar Parameters for J\,1 and J\,37}
\scriptsize
\tablewidth{0pt}
\tablehead{
\colhead{Star} &
\colhead{M$_{\rm V}$\tablenotemark{a}} & 
\colhead{T$_{\rm eff}$(spec)} & 
\colhead{T$_{\rm eff}$(phot)\tablenotemark{b}} & 
\colhead{T$_{\rm eff}$(avg)} & 
\colhead{$\xi_{\rm t}$} &
\colhead{$\log g_{\rm spec}$} & 
\colhead{$\log g_{\rm evol}$\tablenotemark{c}} & 
\colhead{Mass\tablenotemark{c}}\\
\colhead{} & \colhead{} & \colhead{(K)} & \colhead{(K)} & \colhead{(K)} &
\colhead{(km~s$^{\rm -1}$)} & \colhead{} & \colhead{} &
\colhead{(M$_{\odot}$)} 
}

\startdata
J\,1 & $3.47 \pm 0.20$ & $6950 \pm 223$ & $6828 \pm 103$ & $6849 
\pm 93$ & $2.39 \pm 0.18$ & $4.36 \pm 0.23$ & $4.31 
\pm 0.10$ & $1.31 \pm 0.05$\nl
J\,37 & $3.15 \pm 0.20$ & $7305 \pm 242$ & $7122 \pm 93$ & $7145 
\pm 87$ & $3.43 \pm 0.13$ & $3.53 \pm 0.33$ & $\sim 
4.28$ & $\sim 1.45$\nl
\enddata

\tablenotetext{a}{Calculated from an NGC 6633 distance modulus of 7.77 
(Jeffries et al. 2002).}
\tablenotetext{b}{Using the formula given in DSJ02 and $B-V$ colors of
Jeffries (1997).}
\tablenotetext{c}{Derived from 600 Myr Padova Stellar Isochrones (Jeffries 
et al. 2002; Salasnich et al. 2000).}
\end{deluxetable}

\begin{deluxetable}{lcccccc}
\tablecaption{[X/Fe]}
\scriptsize
\tablewidth{0pt}
\tablehead{
\colhead{Element} &
\colhead{J\,37 - J\,1} &
\colhead{Earth\tablenotemark{a}} &
\colhead{mean $\lambda$ Boo\tablenotemark{b}} & 
\colhead{63 Tau\tablenotemark{c}} & 
\colhead{Diffusion model\tablenotemark{c}} & 
}

\startdata
Li & $+0.35$ & $-0.18$ & \nodata & $-0.79$ & $-1.30$ \nl
C & $-1.51$ & $-4.21$ & $+1.00$ & \nodata & $-1.42$ \nl
O & $-1.27$ & $-0.79$ & $+0.70$ & $-1.34$ & $-1.14$ \nl
Na & $-0.10$ & $-0.56$ & $+1.20$ & \nodata & $-0.70$ \nl
Mg & $-0.13$ & $-0.05$ & $+0.00$ & $-0.67$ & $-0.81$ \nl
Al & $-0.04$ & $+0.00$ & $-0.50$ & $-0.44$ & $-0.45$ \nl
Si & $-0.02$ & $-0.09$ & $+1.20$ & $-0.44$ & $-0.85$ \nl
S & $-0.48$ & $-1.14$ & $+0.70$ & \nodata & $-1.12$ \nl
Ca & $-0.06$ & $+0.00$ & $+0.00$ & $-1.31$ & $-1.06$ \nl
Sc & $-0.14$ & $-0.04$ & $-0.10$ & \nodata & \nodata \nl
Ti & $-0.05$ & $-0.05$ & $+0.10$ & \nodata & $-0.45$ \nl
Ni & $+0.18$ & $-0.04$ & $+0.30$ & $+0.31$ & $+0.50$ \nl
\enddata

\tablenotetext{a}{See Gonzalez et al. (2001).}
\tablenotetext{b}{From Heiter (2002).}
\tablenotetext{c}{From RMT00.}
\end{deluxetable}

\end{document}